\documentclass[nofootinbib,aps,preprint,superscriptaddress]{revtex4}
\usepackage{epsfig}
\usepackage{amsmath}

\newcommand{\n}{\overline{n}}
\newcommand{\hs}{\hat{s}}

\newcommand{\mO}{\mathcal{O}}
\newcommand{\mP}{\mathcal{P}}

\newcommand{\mB}{\mathcal{B}}
\newcommand{\mL}{\mathcal{L}}

\newcommand{\mW}{\mathcal{W}}
\newcommand{\mY}{\mathcal{Y}}

\newcommand{\veps}{\varepsilon}
\newcommand{\eps}{\epsilon}

\newcommand{\bea}{\begin{eqnarray}}
\newcommand{\eea}{\end{eqnarray}}
\begin{document}

\baselineskip 3.0ex 

\vspace*{18pt}

\title{Factorization and resummation for single color-octet scalar production 
at the LHC}

\def\addDuke{Department of Physics, Duke University, Durham NC 27708, USA\vspace{0.5cm}} 

\author{Ahmad Idilbi}\email{idilbi@phy.duke.edu}\affiliation{\addDuke}
\author{Chul Kim}\email{chul@phy.duke.edu}\affiliation{\addDuke}
\author{Thomas Mehen}\email{mehen@phy.duke.edu}\affiliation{\addDuke}

\begin{abstract} 
\baselineskip 3.0ex  \vspace{0.5cm}  

Heavy colored scalar particles appear in a variety of new physics (NP) models and  could be
produced at the Large Hadron Collider (LHC). Knowing the total production  cross section is
important for searching for these states and establishing bounds on their masses and couplings. 
Using soft-collinear effective theory, we derive a factorization theorem for the process $pp\to
S X$, where $S$ is a color-octet scalar, that is applicable to any NP model provided the dominant production 
mechanism is gluon-gluon fusion. The factorized result for the inclusive cross section is
similar to that for the Standard Model Higgs production, however, differences arise due to color
exchange between initial and final states. We provide formulae for the total cross section with 
large (partonic) threshold logarithms resummed to next-to-leading logarithm (NLL) accuracy. The 
resulting $K$-factors are similar to those found in Higgs production. We apply our formalism
to the  Manohar-Wise model and find that  the NLL cross section is roughly 2 times (3 times) 
as large as the leading order cross section for a color-octet scalar of mass of 500 GeV (3 TeV). A similar enhancement should appear in any NP model with 
color-octet scalars.

\end{abstract}

\maketitle 
Discovering the Higgs particle and the mechanism of electroweak symmetry breaking is one of the major goals of the Large Hadron Collider (LHC). It is well-known that the main Higgs production mechanism is the gluon-gluon fusion process. An important issue in determining the total production cross section is the large perturbative corrections in the threshold region, defined by $z \to 1$, where $z =m_H^2/\hat{s}$, where $m_H$ is the Higgs mass and $\hat{s}$ is the partonic center of mass energy squared. The leading corrections are enhanced by factors of 
${\rm log}(1-z)/(1-z)$ and invalidate fixed order perturbation theory in the threshold region. 
These corrections can significantly affect the normalization of the total cross section even though
the total cross section receives contributions from a range of $\hat{s}$~\cite{Idilbi:2006dg,
Becher:2007ty,Ahrens:2008nc,Ahrens:2008qu}. In the threshold region, the inclusive scattering cross section $\sigma(p p \to H X )$ can be factorized (at leading twist) into a hard part, soft part, and parton distribution function (PDF) of gluons inside the proton~\cite{Kramer:1996iq}, and the renormalization group equations (RGE) for these parts can be used to resum the large threhold corrections. For the Higgs production cross section the calculations up to the next-to-next-leading logarithm (NNLL) accuracy have already been performed \cite{Kramer:1996iq,Catani:2003zt} and give total cross section about three times bigger than predicted at leading order.

Obviously, properly incorporating these effects will be important for other heavy particles 
predicted in theories of New Physics (NP) that may be observed at the LHC. In this paper 
we will focus on the production of a heavy color-octet scalar, which appears in a number 
of NP models such as grand unified theories~\cite{Dorsner:2007fy,Perez:2008ry,FileviezPerez:2008ib}, 
supersymmetric theories~\cite{Plehn:2008ae,Choi:2008ub},
Pati-Salam unification~\cite{Povarov:2007nh,Popov:2005wz}, chiral color~\cite{Frampton:1987ut}, and
topcolor~\cite{Hill:1991at}. We will calculate the cross section for the Manohar-Wise
 model~\cite{Manohar:2006ga} of color-octet scalars which is consistent with the principle of Minimal Flavor Violation 
(MFV)~\cite{Chivukula:1987py,D'Ambrosio:2002ex}. Performing the resummation
for color-octet scalars is very similar to the resummation for Higgs~\cite{Idilbi:2006dg,Ahrens:2008nc,
Ahrens:2008qu}. Similar resummations have been performed for squark-antisquark and
gluino-pair  production cross sections in Refs.~ \cite{Kulesza:2008jb,Langenfeld:2009eg}.
We will use Soft-Collinear Effective Theory (SCET)~\cite{SCET1,SCETf,Bauer:2002nz} to derive a factorization theorem for the cross section. Gluon-gluon 
fusion cross sections in the full theory are matched onto SCET operators. The matching coefficients
for these operators will differ between various models, but the structure of these operators 
is universal. The cross section computed with the SCET operators factors into correlation
functions of SCET collinear and soft fields, and renormalization group equations
for these correlation functions can be used to perform the resummation of the threshold logarithms. 
This resummation procedure is independent of the NP model.

Before we discuss the factorization theorem,  we will describe some details of the color-octet scalar model we will be focusing on. The principle of MFV requires that the Yukawa couplings of the color-octet scalars be
proportional to the Yukawa matrices in the Standard Model. The Standard Model Yukawa couplings are
\bea
\label{SMyukawa}
{\cal L} = - \, g_{ij}^D \bar{d}_{R i} H^\dagger Q_{L j}
 - \, g_{ij}^U \bar{u}_{R i} \epsilon H^* Q_{L j} + h.c. \, ,  
\eea
where $Q_L$ is the doublet of lefthanded quarks, $u_R$ and $d_R$ are the righthanded up and down 
quarks, respectively, $i$ and $j$ are flavor indices, and $H$ is the Higgs doublet:
\bea
H =\left(\begin{array}{c} H^+ \\ H^0 \end{array}\right) \qquad \epsilon H^* = \left(\begin{array}{c} H^{0*} \\ -H^- \end{array}\right) \, .
\eea 
If the color-octet scalar Yukawa couplings are  
\bea
\label{yukawa}
{\cal L} = - \, \lambda_{ij}^D \bar{d}_{R i} S^{a \dagger} T^a Q_{L j} - \, 
\lambda_{ij}^U \bar{u}_{R i} \epsilon  S^{a *} T^a Q_{L j}\, , + h.c. \, .
\eea
then the MFV hypothesis requires  $\lambda_{ij}^{U,D}= \eta_{U,D}\,g^{U,D}_{ij}$, where $\eta_{U,D}$ are constants.  This eliminates tree-level flavor changing neutral currents and ensures that experimental constraints from flavor physics are not violated. Note that this also implies that the $S^a$ couple most strongly to the third generation of quarks. The color-octet scalars have gauge couplings to gluons and a gauge
invariant mass term
\bea
 \label{fullla} 
\mL_{\mathrm{QCD}} = - \frac{1}{2} S^a (D^2)^{ac} S^c - \frac{1}{2} m_S^2 S^a S^a \, ,  
\eea
where $D_{\mu}^{ac} = \partial_{\mu} \delta^{ac} + g f^{abc}A_{\mu}^b$. Finally, there is a scalar potential for the $S^a$ and the Higgs
that can be found in Ref.~\cite{Manohar:2006ga}.

Though it may seem ad hoc to impose $\lambda_{ij}^{U,D} \propto \,g^{U,D}_{ij}$, this can arise naturally 
in certain models. For example, consider the chiral color model of Ref.~\cite{Frampton:1987ut}. In this model, the gauge 
group is enlarged to $SU(3)_L \times SU(3)_R \times SU(2)_L \times U(1)_Y$, and the chiral color group
$SU(3)_L \times SU(3)_R$ breaks down to $SU(3)_c$ at some high energy scale. If the righthanded quarks
are placed in the $(1,3,1)$ representation of $SU(3)_L \times SU(3)_R \times SU(2)_L$, and the lefthanded 
quarks are placed in the $(3,1,2)$, then quarks can obtain masses from the following Yukawa couplings
\bea
{\cal L} = - \, \sqrt{3} \,g_{ij}^D \bar{d}_{R i} \Phi^\dagger Q_{L j}
 - \, \sqrt{3} \, g_{ij}^U \bar{u}_{R i} \epsilon \Phi^{\prime *} Q_{L j} + h.c. \, ,  
\eea
where $\Phi$ and $\Phi^\prime$ transform in the  $(3,\bar{3}, 2)$ and $(\bar{3},3,2)$, respectively.\footnote{Note that an additional scalar in the $(1,1,2)$ is required to give masses to leptons.}
%Upon the symmetry breaking $SU(3)_L \times SU(3)_R  \rightarrow SU(3)_c$ 
The fields $\Phi$, $\Phi^{\prime}$ can be decomposed into singlet and octet scalars under the unbroken $SU(3)_c$. 
\bea 
\Phi^{(\prime)} = \frac{1}{\sqrt{3}} H^{(\prime)} + S^{a (\prime)} T^a 
\eea
and $\lambda_{ij}^{U,D} =\sqrt{3} \,g^{U,D}_{ij}$ at tree level. Note that this model is different
from the Manohar-Wise model since there are two distinct color-octet scalars, $S^{a}$ and $S^{a \prime}$. If the gluon-gluon fusion production of a single 
color-octet scalar proceeds through a top-quark loop, this mechanism will predominantly produce 
$S^{a \prime}$. We do not wish to to study this model further, but mention it to show that 
the MFV constraint $\lambda_{ij}^{U,D} \propto \,g^{U,D}_{ij}$ could emerge as a consequence of a symmetry of the underlying theory. 

% is inspired by some models of ``beyond the Standard Model'' (BSM). Such models allow an extension of the gauge 
%group $\mathrm{SU}_{\mathrm{C}} (3) \times \mathrm{SU}_{\mathrm{L}} (2)$ to the (8,2) representation \cite
%{Manohar:2006ga} or the breakdown of $\mathrm{SU}_{\mathrm{C}} (3) \times \mathrm{SU}_{\mathrm{HC}} (3) \to 
%\mathrm{SU}_{\mathrm{C}} (3)$ \cite{Dobrescu:2007yp,Kilic:2008pm}, where $\mathrm{SU}_{\mathrm{HC}} (3)$ is a 
%hidden color gauge symmetry. 

The color-octet scalars can be produced via the pair production cross section, $gg \to S S$ which proceeds through
gauge couplings~\cite{Manohar:2006ga}. Constraints on pair production followed by decay to heavy quarks has been used to
establish a lower bound on $m_S$ of about $200$ GeV in Ref.~\cite{Gerbush:2007fe}.
Neutral color-octet scalars can also be produced singly via gluon-gluon fusion~\cite{Gresham:2007ri}. 
This proceeds through loop diagrams containing quarks, of which the top quark gives by far the dominant contribution,  and loops with color-octet scalars. The relative size of the top quark and  scalar loop contributions
is determined by  $\eta_U$ and other parameters in the scalar potential. If these parameters are all taken to be of order
unity then the top quark loop is the largest contribution.~\footnote{ If the color-octet is a pseudoscalar only the top quark loop contributes.}
In this case the production mechanism is very similar to that for a single Higgs, and hence threshold corrections are expected to be significant. At the LHC 
the production cross section for single $S$ production is larger than pair production when the mass of $S$ is larger than 1 TeV~\cite{Gresham:2007ri}.  

%In this work we are interested, for phenomenological studies, to improve the LO results of \cite{Manohar:2006ga} by obtaining a NLL resummed cross section. 
%This could only be achieved after a proper factorization theorem is proved and the resummation procedure of threshold logarithms will be performed by 
%adopting the effective field theory approach. As we mentioned earlier, in the partonic threshold, the only relevant momentum modes are soft and collinear 
%ones. Thus the relevant effective theory is the soft collinear effective theory (SCET) \cite{SCET1,SCETf}. %Because SCET has various field contents 
%according to the kinematic features which do not interact with each other, the factorization theorem can be intrinsically employed. After matching, SCET 
%reproduces the low energy physics in full QCD and the Wilson coefficients of the matched SCET operators provide the perturbative results of the electroweak
%dynamics and the hard strong interactions. So the short distance expansion is properly performed reflecting the systematic power expansion of the small 
%$\lambda$ in SCET. Also the solutions of the Wilson coefficients for the renormalization group equations (RGE) can be automatically identified with the
% resumed result of large Sudakov logarithms.  

When applying SCET to  color-octet scalar production we first match full QCD onto SCET operators at the hard scale $\mu_h \sim m_S$, where $m_S$ is the mass of the color-octet field. SCET is formulated as an expansion 
in $\lambda \sim \sqrt{\mu_s/\mu_h}$, where  $\mu_s$ is the soft scale, $\mu_s \sim m_S (1-z)$. 
The allowed SCET operators are constrained by SCET gauge symmetries~\cite{SCETgauge}. 
At leading order in $\lambda$, we find two  dimension-5 operators  with different color structures. Subleading operators will be constrained by the requirement of reparametrization invariance \cite{RPI}. 

%The important point we like to emphasize is that the operator basis in SCET can be determined irrespective of any NP model. In this sense it is universal. 
%However the Wilson coefficient obtained at the first step matching is dependent on the specifics of the NP model. Below the hard scale, the surviving strong
%interactions are mainly either soft or collinear modes which describe the soft gluon radiations and gluon PDFs respectively. 
%Therefore we note that all the dynamics below the hard scale near $m_S$ are independent of the NP model and can be completely encoded via the SCET operators.
%As we show below, that even though the color-octet particle is strongly interacting in contrast to Higgs particle, the (gauge invariant) strong interaction
%Lagrangian for the color-octet can be set up in a unique manner and it guarantees the universality of the QCD factorization theorem on the single color-octet
%production. 

Our result for  the factorized scattering cross section is
\begin{eqnarray} 
\label{sca}
\sigma (pp\to S X) &=& \tau H(m_S,\mu_f) \int^1_{\tau} \frac{dz}{z} \bar{S}(m_S(1-z),\mu_f)  F(\tau/z,\mu_f).  
\end{eqnarray} 
Here $\tau = m_S^2/s$ where $s$ is the center of mass energy squared at the LHC, $H(m_S,\mu)$ and $\bar{S}(m_S(1-z),\mu)$ are the hard and soft functions respectively, and $F(x,\mu)$ is the following convolution of the gluon PDF's: 
\begin{equation} 
\label{lumi} 
F(x,\mu_f) =\int^1_{x} \frac{dy}{y} f_{g/P} (y,\mu_f) f_{g/P} (x/y,\mu_f).
\end{equation} 
%The sum over $i$ corresponds to the two previously mentioned operators, which we label $f$ and $d$ for reasons that will be clear below. 
Renormalization group equations for the $H(m_S, \mu_f)$,
$\bar{S}(m_S(1-z),\mu_f)$, and $F(\tau/z,\mu_f)$ can be used to resum large threshold corrections as will
be discussed below.

%In Eq.~(\ref{sca}), the hard functions, $H_i(M_s,\mu_f)$ are obtained perturbatively from the matching %at the scale $\mu_h \sim m_S$ and then are evolved down to $\mu_s\sim m_S (1-z)$, which is the
 %typical scale of the soft gluon radiation. Since the scale $\mu_s$ is  larger than 
 %$\Lambda_{\rm QCD}$, we then integrate out the radiated soft gluons and then obtain the soft %functions $S_i(1-z,\mu_f)$. 
%Finally we evolve the soft functions from $\mu_s$ to some arbitrary factorization scale, $\mu_f$, and %the gluon PDFs given at the scale $\mu_f$ describe the remaining collinear degrees of freedom in the %scattering process. 

In order to prove the factorization theorem in Eq.~(\ref{sca}) we need to construct the SCET operators composed of the collinear  gluons from the initial state hadrons, soft  gluons, and a heavy color-octet scalar field. In the center of the mass frame, the incoming gluons from the two incoming protons are described as $n$ and $\n$-collinear fields where the light-cone vectors satisfy $n^2=\n^2=0,~n\cdot \n=2$. The lowest dimension operator with a single $n$-collinear gluon that is $n$-collinear
gauge invariant is $W_n^{\dagger} G_n^{\mu\nu} W_n$. Here $G_n^{\mu\nu} = G_n^{a,\mu\nu} T^a$ is a SCET gluon field strength tensor, and $W_n$ is the collinear Wilson line 
\begin{equation} 
\label{cw} 
W_n (x) = \mathrm{P} \exp \Biggl( ig \int^x_{-\infty} ds ~\n\cdot A_n^a (s\n^{\mu}) T^a \Biggr).
\end{equation}
Similarly,  the lowest dimension $\n$-collinear gauge invariant operator 
 is  $W_{\n}^{\dagger} G_{\n}^{\mu\nu} W_{\n}$. 
Combining  $W_n^{\dagger} G_n^{\mu\nu} W_n$ and $W_{\n}^{\dagger} G_{\n}^{\mu\nu} W_{\n}$ into a Lorentz scalar 
and then expanding to lowest order in $\lambda$ yields \begin{eqnarray} 
\label{gc} 
\Bigl(W_n^{\dagger} G_n^{\mu\nu} W_n\Bigr)_{\alpha\beta} \Bigl(W_{\n}^{\dagger} G_{\n,\mu\nu} W_{\n}\Bigr)_{\gamma\delta} &=& -\frac{1}{g^2} \Bigl(W_n^{\dagger} [iD_n^{\mu},iD_n^{\nu}]  W_n\Bigr)_{\alpha\beta} \Bigl(W_{\n}^{\dagger} [iD_{\n\mu},iD_{\n\nu}] W_{\n}\Bigr)_{\gamma\delta} \\ 
&=& -\Bigl(\mB_n^{\perp\mu}\Bigr)_{\alpha\beta} \Bigl(\mB_{\n,\mu}^{\perp}\Bigr)_{\gamma\delta} + \mO (\lambda) \nonumber.  
\end{eqnarray} 
Here $\alpha, \beta, \gamma$, and $\delta$ are color indices in the fundamental representation and 
$\mB_n^{\perp\mu}$ is 
\begin{equation} 
\mB_n^{\perp\mu} = \frac{1}{g} \Bigl[\n\cdot \mP W_n^{\dagger} iD_n^{\perp\mu} W_n \Bigr],
\end{equation} 
where the derivative operator $\mP^{\mu}$ returns the large label momentum and only  acts on collinear fields within the brackets $[\cdots]$.
It will be convenient to write the field $\mB_n^{\perp \mu}$ in terms of the Wilson line in the adjoint representation (i.e. with color generator $(t^a)_{bc} = -if^{abc}$). Defining $\mB_n^{\perp\mu}=\mB_{n\perp}^{a,\mu} T^a$,  $\mB_{n\perp}^{a,\mu}$ is given by   
\begin{equation} 
\label{adjointB} 
\mB_{n\perp}^{a,\mu} = i\n^{\rho} g_{\perp}^{\mu\nu} \mW_n^{\dagger,ab} G_{n,\rho\nu}^b
=i\n^{\rho} g_{\perp}^{\mu\nu} G_{n,\rho\nu}^b \mW_n^{ba},
\end{equation} 
where $\mW_n$ is the collinear Wilson line in the adjoint representation, and we used the relation $\mW_n^{ab} = \mW_n^{\dagger ba}$ for the second equality. For the $\n$-collinear fields,  
 $\mB_{\n}^{\perp\mu}$ and $\mB_{\n\perp}^{a,\mu}$ are identical to $\mB_n^{\perp\mu}$ and $\mB_{n\perp}^{a,\mu}$, respectively, 
after interchanging $n$ and $\n$.

Finally, we need to include fields for the color-octet scalar. The strong interactions of this field are described by Eq.~(\ref{fullla}).
At scales well below $m_S$, the strong interactions of the heavy color-octet scalar simplify because the scalar is slowly moving. In the threshold region,
the $S^a$ is produced nearly at rest (in the parton center-of-mass frame) and heavy particle effective theory techniques can be applied.
We use a heavy scalar effective theory (HSET), similar to heavy quark effective theory (HQET). In HSET, the scalar momentum is decomposed
into large and small parts: $p^\mu_S = m_S v^\mu + k^\mu$, where $v^\mu$ is the static four-velocity and $k^\mu$ represents fluctuations   of $O(\mu_s)$. In order for derivatives in HSET to bring factors of $k^\mu$ rather than the total momentum, we use the standard rephasing trick to relate full theory and HSET fields :
\begin{equation} 
\label{phifull} 
S^a (x) = \frac{1}{\sqrt{2m_S}}\Bigl(e^{-im_Sv\cdot x} S_v^a (x) + e^{im_Sv\cdot x} S_v^{*a} (x)\Bigr). 
\end{equation} 
The HSET Lagrangian is obtained by
plugging Eq.~(\ref{phifull}) into Eq.~(\ref{fullla}) and taking the large $m_S$ limit, in which
the only surviving terms are those  for which the phase factor cancels. We find 
\begin{equation} 
\label{HSET} 
\mL_{\mathrm{HSET}} = S_v^{*a} (v\cdot iD_s)^{ac} S_v^c - \frac{1}{2m_S} S_v^{*a} (D_s^2)^{ac} S_v^c,
\end{equation} 
where $v^\mu$ is the four-velocity and the covariant derivative, $D^\mu_s$, involves only soft gluons. The first term in Eq.~(\ref{HSET}) gives the leading interactions, and the second term is suppressed by $1/m_S$ and so can be neglected. 

In our SCET-HSET operators, the  soft gluons appear in the soft Wilson lines, 
\begin{equation} 
\label{sw} 
\mY_{\mathrm{v}} (x) = \mathrm{P} \exp \Biggl( ig \int^x_{-\infty} ds ~\mathrm{v}\cdot A_s^a (s \mathrm{v}^{\mu}) t^a \Biggr),
\end{equation}
where $\rm{v}^{\mu}$ can be either $n^\mu$, $\n^\mu$, or $v^\mu$. These soft Wilson lines arise  when we decouple the leading soft interactions from the collinear and heavy scalar fields by the field redefinitions \cite{SCETf}
\begin{eqnarray} 
\label{decouple} 
&&A_n^{a,\mu} \to \mY_n^{ab} A_n^{b,\mu},~~~\mW_n^{(\dagger)ab} \to \mY_n^{ac}\mW_n^{(\dagger)cb} \\ 
&&A_{\n}^{a,\mu} \to \mY_{\n}^{ab} A_{\n}^{b,\mu},~~~\mW_{\n}^{(\dagger)ab} \to \mY_{\n}^{ac}\mW_{\n}^{(\dagger)cb} \nonumber \\
&&S_{v}^{a} \to \mY_{v}^{ab} S_{v}^{b},~~~S_{v}^{*a} \to \mY_{v}^{ba} S_{v}^{*b}. \nonumber 
\end{eqnarray}
After this field redefinition, the  collinear fields and HSET fields do not interact with the soft particles.
Note that after the field redefinition, $\mL_{\rm{HSET}} =S_v^{a*} (v\cdot i\partial) S^a_v + O(1/m_S)$, so the strong interactions vanish at leading order in $O(1/m_S)$. The heavy scalar's interaction with soft gluons can be reproduced by a soft Wilson line. This replacement simplifies the derivation of the factorization theorem as we show below.

Using $\mB_{n}^{\perp\mu}$, $\mB_{\n}^{\perp\mu}$, $S^a_v$, and the soft Wilson lines 
we can construct the effective Lagrangian for color-octet production at leading order in $\lambda$,
\begin{equation} 
\label{SCET} 
\mL_{\mathrm{SCET}} = C_S(\mu)O_S(\mu) + C_P(\mu)O_P(\mu),
\end{equation}
where the effective theory operators $O_S$ and $O_P$ are for color-octet scalars ($S_{S}$) and pseudoscalars ($S_{P}$) respectively.
Those operators have different color structures and are 
\begin{eqnarray} 
\label{Of}
O_S &=& \frac{d^{abc}}{\sqrt{2m_S}} (S_{Sv}^* \mY_v^{\dagger} )^a (\mB_{\n}^{\perp\mu} \mY_{\n}^{\dagger})^b (\mY_{n} \mB_{n\mu}^{\perp} )^c, \\
\label{Od}
O_P &=& \frac{if^{abc}}{\sqrt{2m_S}} \eps^{\perp}_{\mu\nu} (S_{Pv}^* \mY_v^{\dagger} )^a (\mB_{\n}^{\perp\mu} \mY_{\n}^{\dagger})^b (\mY_{n} \mB_{n}^{\perp\nu} )^c,
\end{eqnarray}
where $\eps^{\perp}_{\mu\nu} = \eps_{\mu\nu\rho\sigma}n^{\rho}\n^{\sigma}/2$.
%Since the $n$-collinear, $\n$-collinear, and the soft fields are decoupled, the renormalization of these operators will 
%be a simple product given by
%\bea
%Z_{f,d} = Z_{\mY}\, Z_n \, Z_{\n} \,, 
%\eea
%where $Z_n$ and $Z_{\n}$ are the renormalization factors for the collinear parts $\mB_{n}^{\perp \mu}$ and $\mB_{\n}^{\perp\mu}$,
%respectively, and $Z_{\mY}$ is the renormalization factor for the three soft Wilson lines. The two soft Wilson line operators,
%$f^{abc} {\cal Y}^{\dagger a^\prime a} {\cal Y}^{\dagger b^\prime b}{\cal Y}^{\dagger c\prime c}$
% and $d^{abc} {\cal Y}^{\dagger a^\prime a} {\cal Y}^{\dagger b^\prime b}{\cal Y}^{\dagger c\prime c}$,
%are antisymmetric and symmetric, respectively,  under the exchange of any pair of color indices: $a^\prime, b^\prime$ and $c^\prime$, and therefore these operators do not mix under renormalization to all orders in perturbation theory.
%If the color-octet particle were a pseudoscalar, then a similar analysis leads to the following SCET-HSET operators:
%\begin{eqnarray} 
%\label{Opf}
%O_f^p &=& \frac{if^{abc}}{\sqrt{2m_S}} \eps^{\perp}_{\mu\nu} (S_v^* \mY_v^{\dagger} )^a (\mB_{\n}^{\perp\mu} \mY_{\n}^{\dagger})^b (\mY_{n} \mB_{n}^{\perp\nu} )^c, \\
%\label{Opd}
%O_d^p &=& \frac{d^{abc}}{\sqrt{2m_S}} \eps^{\perp}_{\mu\nu} (S_v^* \mY_v^{\dagger} )^a (\mB_{\n}^{\perp\mu} \mY_{\n}^{\dagger})^b (\mY_{n} \mB_{n}^{\perp\nu} )^c,
%\end{eqnarray} 
%where $\eps^{\perp}_{\mu\nu} = \eps_{\mu\nu\rho\sigma}n^{\rho}\n^{\sigma}/2$. 
Note that the strong interaction Lagrangian for the pseudoscalar is the same as the scalar. 
Since the $n$-collinear, $\n$-collinear, and the soft fields are decoupled, the renormalization of the both operators will be a simple product given by
\bea
Z_{S,P} = Z_{\mY}\, Z_n \, Z_{\n} \,, 
\eea
where $Z_n$ and $Z_{\n}$ are the renormalization factors for the collinear parts $\mB_{n}^{\perp \mu}$ and $\mB_{\n}^{\perp\mu}$, respectively, and 
$Z_{\mY}$ is the renormalization factor for the three soft Wilson lines. 
So the renormalizations of $O_S$ and $O_P$ are the same and do not depend on either color structure constants or the Lorentz structure. 
%as for the scalar case. 
%operators in Eqs.~(\ref{Of},\ref{Od}).
Therefore, below we only consider the renormalization of the  scalar operators, however our results hold for  pseudoscalar operators as well. 

Taking the matrix elements of $C_S(\mu)O_S(\mu)$ in Eq.~(\ref{SCET}), 
we find the scattering cross section 
\begin{eqnarray} 
\label{scaS1} 
\sigma (pp\to S_S X) &=& 
\frac{\pi}{s} \sum_X \int d^4 q \delta (q^2-m_S^2) \delta (P_n+P_{\n} -q -p_X) \nonumber \\ 
&&\times |C_S(m_S, \mu)|^2 \langle P_nP_{\n} | O_{S}^{\dagger} (\mu) | S_S X \rangle \langle S_S X | O_{S} (\mu) | P_nP_{\n} \rangle, \nonumber 
\end{eqnarray} 
where $P_n^\mu$ and $P_{\n}^\mu$ are the momenta of the incoming protons which are $n$-collinear and  $\n$-collinear, respectively. Because the final state $X$ consists of $n(\n)$-collinear and soft states in the partonic threshold region, it is possible to rewrite the final state summation as $\sum_X = \sum_{X_n}\sum_{X_{\n}}\sum_{X_S}$ and the final state momentum $p_X = p_{X_n} + p_{X_{\n}}+p_{X_S}$. The momentum of the color-octet field is 
\begin{equation} 
\label{higgsmom} 
q = P_n+P_{\n} - (p_{X_n} + p_{X_{\n}}+p_{X_S}) = p_n+p_{\n} -p_{X_S},
\end{equation} 
where $p_n = P_n-p_{X_n}$ and $p_{\n} = P_{\n}-p_{X_{\n}}$ are momenta of the partons in the two incoming protons. Then the argument in the first delta function in Eq.~(\ref{scaS1}) is 
\begin{eqnarray} 
\label{higgsmomsq} 
q^2-m_S^2 &=& (p_n+p_{\n} -p_{X_S})^2 -m_S^2 \sim (p_n+p_{\n})^2 -2p_{X_S} \cdot (p_n+p_{\n}) -m_S^2\\
&=&\hat{s} -2\eta\hat{s}^{1/2} -m_S^2, \nonumber 
\end{eqnarray} 
where $\eta = p_{X_S}^0$ is the energy carried by final state soft particles in the partonic center-of-mass frame, and  $\hat{s} = \n\cdot p_n n\cdot p_{\n} = y_1 y_2 s$, where $y_{1,2}$ are the large (collinear) momentum fractions of the incoming partons, defined by $y_1=\n\cdot p_n /\n\cdot P_n$ and $y_2=n\cdot p_{\n} /n\cdot P_{\n}$.
When the momentum $q$ is on-shell,  Eq.~(\ref{higgsmomsq}) implies 
\begin{equation} 
\label{etaeq} 
\eta = \frac{\hs-m_S^2}{2\hs} = \frac{\hs (1-z)}{2} = \frac{m_S(1-z)}{2z^{1/2}} \sim \frac{m_S(1-z)}{2},
\end{equation} 
where the last approximation is valid in the limit $z=m_S^2/\hs = \tau/(y_1y_2) \to 1$.

 The cross section in Eq.~(\ref{scaS1}) can be factorized by first inserting the identity
\bea\label{id}
1=\int d\eta dy_1 dy_2 \delta(\eta+i\partial_0)\delta\left(y_1 -\frac{\n\cdot \mP}{\n\cdot P_n}\right)
\delta\left(y_2-\frac{n\cdot \mP}{n\cdot P_{\n}}\right),
\eea
where $\n \cdot \mP$ is a label operator acting on $n$-collinear fields, 
$n \cdot \mP$ is a label operator acting on $\n$-collinear fields, and the partial 
derivative, $i \partial_0$, acts only on soft fields. Then $\sigma(pp\to S_SX)$ can be written as
\begin{eqnarray} 
\sigma(pp\to S_S X) &=& \frac{\pi}{s} \int d\eta dy_1 dy_2 \delta\Bigl(\eta -\frac{m_S(1-z)}{2} \Bigr) \frac{|C_S(m_S, \mu)|^2}{2\hs^{1/2}} \\
&&\times \frac{d^{abc}d^{def}}{2m_S}\Bigl\langle P_nP_{\n}  \Bigl| 
(\mY_vS_{Sv})^a (\mB_{n}^{\perp\mu} \mY_{n}^{\dagger})^b (\mY_{\n} \mB_{\n\mu}^{\perp} )^c \Bigr| S_S X \Bigr\rangle \nonumber \\
&&\times \Bigl\langle S_S X \Bigl| \delta(\eta+i\partial_0) 
(S_{Sv}^* \mY_v^{\dagger} )^d (\mB_{\n\mu}^{\perp}[y_2] \mY_{\n}^{\dagger})^e 
(\mY_{n} \mB_{n\mu}^{\perp}[y_1] )^f \Bigr| P_nP_{\n} \Bigr\rangle.  \nonumber 
\eea
The cross section $\sigma(pp\to S_P X)$ is the same up to the color factor and the replacement $C_S(m_S,\mu)\to C_P(m_S,\mu)$.  Then we use 
$S_v^a | S \rangle = \sqrt{2m_S} \epsilon^a$ and
 $\langle S | S_v^{*b} = \sqrt{2m_S} \epsilon^{*b}$
to remove the color-octet scalar from the final state, where $\epsilon^a$ are color polarization vectors satisfying the relations, $\sum_{pol}\epsilon^{*a} \epsilon^{b}=\delta^{ab}$. 
Finally, we apply the completeness relation, $1=\sum_X |X \rangle \langle X |$.
After these manipulations, we find that 
$\sigma(pp\to S_S X)$ is given by
\bea \label{scaf}
\sigma(pp\to S_S X) 
&=& \frac{\pi d^{abc}d^{def} }{s } \int dy_1 dy_2 \frac{|C_S(m_S, \mu)|^2}{2\hs^{1/2}} \Bigl\langle P_nP_{\n} \Bigl| \mY_{v}^{ak} \mB_n^{\perp\mu,l} \mY_n^{\dagger lb} \mY_{\n}^{cm} \mB_{\n\mu}^{\perp,m}  \\
&&\times \delta\Bigl( \frac{m_S(1-z)}{2} + i\partial_0 \Bigr)\mY_v^{\dagger kd} \mB^{\perp,n}_{\n,\nu} [y_2] \mY_{\n}^{\dagger ne} \mY_n^{fo} \mB^{\perp\nu,o}_n [y_1] \Bigr| P_nP_{\n} \Bigr\rangle. \nonumber 
\end{eqnarray} 

To simplify the notation we have defined 
\begin{equation} 
\mB^{\perp\mu,a}_n [y_1] = \Bigl[\delta\Bigl(y_1-\frac{\n\cdot \mP}{\n\cdot P_n}\Bigr)\mB^{\perp\mu,a}_{n}\Bigr],~~~\mB^{\perp\mu,a}_{\n} [y_2] = \Bigl[\delta\Bigl(y_2-\frac{n\cdot \mP}{n\cdot P_{\n}}\Bigr)\mB^{\perp\mu,a}_{\n}\Bigr].
\end{equation}  
The PDF for the $n$-collinear proton in terms of SCET fields is 
\begin{eqnarray} 
\label{SCETPDF} 
f_{g/P_n} (x) &=&  \frac{1}{2\pi x \n\cdot P_n} \int \frac{dn\cdot z}{2} e^{-ix \n\cdot P_n n\cdot z/2}\\
&&\times \n^{\alpha}\n^{\beta} g_{\perp}^{\mu\nu} \Bigl\langle P_n \Bigl| G_{n,\alpha\mu}^a \left(\frac{n\cdot z}{2}\right) \mW_n^{ab} \Bigl[\frac{n\cdot z}{2},0\Bigr] G^b_{n,\beta\nu} (0) \Bigr| P_n\Bigr\rangle \nonumber \\
&=&\frac{1}{x \n\cdot P_n} \n^{\alpha}\n^{\beta}g_{\perp}^{\mu\nu} \Bigl\langle P_n \Bigl| G_{n,\alpha\mu}^a  \mW_n^{ab} \delta(x \n\cdot P_n - \n\cdot \mP) \mW_n^{\dagger bc} G^c_{n,\beta\nu} \Bigr| P_n\Bigr\rangle \nonumber \\
&=& \frac{1}{x(\n\cdot P_n)^2} \Bigl\langle P_n \Bigl| \mB_{n\perp}^{a,\mu}  \delta\Bigl(x  - \frac{\n\cdot \mP}{\n\cdot P_n} \Bigr) \mB_{n\mu}^{\perp,a} \Bigr| P_n\Bigr\rangle \nonumber 
\end{eqnarray} 
where $P_n^{\mu}$ is the momentum of the proton. Here, we have defined $\mW_n^{ab}[z,0] = \mW_n^{ab} (z) \mW_n^{\dagger bc} (0)$ and  used Eq.~(\ref{adjointB}) for the third equality. 
A similar set of manipulations yields the PDF for the $\n$-collinear proton, and since
 $f_{g/P_n}(x) = f_{g/P_{\n}}(x)$, we will drop the subscripts $n$ and $\n$ in what follows.  
Therefore, after averaging over proton spins in Eq.~(\ref{scaf}),
 we find
\begin{eqnarray} 
\label{pdfproject} 
\langle P_n | \mB_n^{\perp\mu,l} \mB_n^{\perp\nu,o} [y_1] | P_n \rangle &=& g_{\perp}^{\mu\nu} \delta^{lo}\frac{y_1 (\n\cdot P_n)^2}{2(N_c^2-1)} f_{g/P} (y_1), \\ 
\langle P_{\n} | \mB_{\n\mu}^{\perp,m} \mB_{\n\nu}^{\perp,n} [y_2] | P_{\n} \rangle &=& g^{\perp}_{\mu\nu} \delta^{mn}\frac{y_2 (n\cdot P_{\n})^2}{2(N_c^2-1)} f_{g/P} (y_2). \nonumber
\end{eqnarray} 
The soft function $\bar{S}_S~(\bar{S}_P)$ for the scalar (pseudoscalar) production is defined to be 
\begin{eqnarray} 
\label{softS} 
\bar{S}_S (m_S(1-z)) &=& \frac{3d^{abc}d^{def}}{40} \Bigl\langle 0 \Bigl| \mY_v^{ak} \mY_n^{bl} \mY_{\n}^{cm} \delta\Bigl(1-z+2\frac{i\partial_0}{m_S} \Bigr) \mY_v^{\dagger kd}\mY_{n}^{\dagger le} \mY_{\n}^{\dagger mf} \Bigr| 0 \Bigr\rangle, \\
\label{softP}
\bar{S}_P (m_S(1-z)) &=& \frac{f^{abc}f^{def}}{24} \Bigl\langle 0 \Bigl| \mY_v^{ak} \mY_n^{bl} \mY_{\n}^{cm} \delta\Bigl(1-z+2\frac{i\partial_0}{m_S} \Bigr) \mY_v^{\dagger kd}\mY_{n}^{\dagger le} \mY_{\n}^{\dagger mf} \Bigr| 0 \Bigr\rangle.
\end{eqnarray} 
Here, the soft functions are normalized to $\delta(1-z)$ at lowest order. 

Using these definitions
we obtain the factorized scattering cross sections which are given by   
\begin{equation} \label{factthm}
\sigma (pp\to S_i X) = 
\tau H_i (m_S,\mu_f)\int^1_{\tau} \frac{dz}{z} \bar{S}_{i}(m_S(1-z),\mu_f) F(\tau/z,\mu_f),~~~i=S,P,
\end{equation} 
where the hard coefficients $H_{i}$ are 
\begin{equation} 
H_S (m_S, \mu_f) = \frac{5\pi |C_S(m_S, \mu_f)|^2}{48},~~~H_P (m_S, \mu_f) = \frac{3\pi |C_P(m_S, \mu_f)|^2}{16}.
\end{equation}
This factorization theorem is our main result. When we consider the renormalization group evolution effects, we will calculate $H_{S,P}(m_S,\mu)~(\bar{S}_{S,P}(m_S(1-z),\mu))$ at the scale $\mu_h~(\mu_s)$ and then evolve them to the factorization scale $\mu_f$. 
At the leading order (LO) in $\alpha_s$, the cross section is 
\bea 
\sigma(pp \to S_i X) =  \tau H_i^{(0)} F(\tau),~~~i=S,P,
\eea
where $H_{i}^{(0)}$ are the hard coefficient at the lowest order and are equal to the scattering cross section at the Born level. 
%\begin{equation} 
%\label{HiggsBorn2} 
%\sigma_0 = \pi\Bigl(\frac{3}{16}|C_f^{(0)}(m_S)|^2+\frac{5}{48}|C_d^{(0)}(m_S)|^2\Bigr).
%\end{equation} 
%This expression allows one to relate the calculation of $\sigma(gg \to S^a)$ in any NP 
%model to the coefficients appearing in the factorization theorem, Eq.~(\ref{scaf}).

\begin{figure}[t]
\begin{center}
\epsfig{file=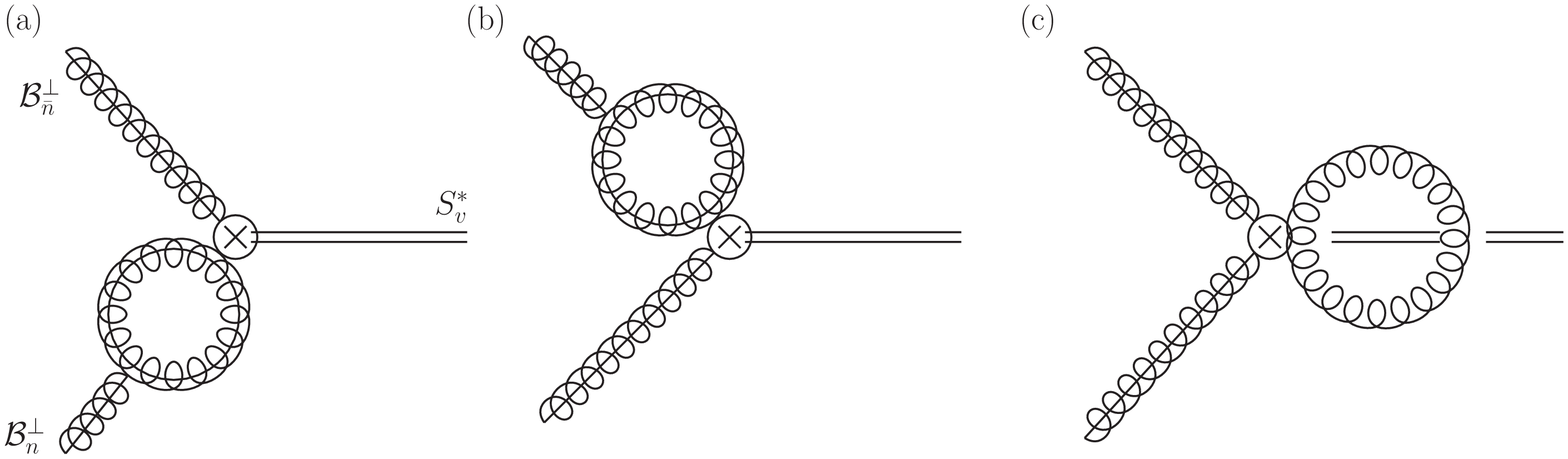, width=15cm, height=4cm}
\end{center}
\vspace{-0.7cm}
\caption{\baselineskip 3.0ex  
 One-loop renormalization of $O_{S,P}$. Here the curly lines with the straight lines are $n(\n)$-collinear gluons and the only curly lines denote the soft gluons coming from the soft Wilson lines. Double line denotes outgoing color-octet field.  
} 
\label{fig1} 
\end{figure}

Next we discuss the RGE evolution of the hard and soft parts  and  the
resummation of the cross section.  Once the evolution
for the coefficient functions, $C_{S,P}(\mu)$,  is determined, the evolution equation for the soft function 
can be easily derived, since the evolution equations for  $f_{g/p}(x)$ are known. 
The evolution of the soft functions can be done in momentum space as in 
the analysis of Higgs production and 
Drell-Yan in Refs.~\cite{Becher:2007ty,Ahrens:2008nc,Ahrens:2008qu}. We follow this
approach in this paper. Alternatively, one can
solve evolution equations for the moments of the soft functions and PDF's, and then take 
an inverse Mellin transform to obtain the resummed cross section. Resummed expressions
for the moments of the cross section are given in the Appendix.

To determine the one-loop anomalous dimensions of $O_{S,P}$, we need to consider the Feynman diagrams in Fig.~\ref{fig1} as well as the wavefunction renormalization graphs. We regulate ultra-violet (UV) divergences using dimensional regularization and the infrared (IR) divergences by taking the external legs to be off-shell. It is then straightforward to extract the UV divergences and we find
\begin{equation} 
\label{ZO} 
Z_S = Z_P = 1+\frac{\alpha_s}{4\pi}\frac{1}{\veps_{\rm{UV}}}\Bigl[N_c \Bigl(\frac{2}{\veps_{\rm{UV}}}+2\ln\frac{\mu^2}{m_S^2} + \frac{14}{3} + i\pi\Bigr)-\frac{2}{3} n_f \Bigr],
\end{equation} 
where $N_c$ and $n_f$ are the number of colors and flavors, respectively. From $Z_{S,P}$ we obtain the anomalous dimension for $O_S$ and $O_P$, 
\begin{equation} 
\label{gammaH}
\gamma_{S,P} =  \Bigl(\mu\frac{\partial}{\partial\mu} + \beta\frac{\partial}{\partial g}\Bigr)\ln Z_{S,P} = -\frac{\alpha_s}{\pi} \Bigl[N_c\Bigr(\ln\frac{\mu^2}{m_S^2}+\frac{7}{3} +i\frac{\pi}{2}\Bigr) - \frac{n_f}{3} \Bigr].
\end{equation} 
Note that 
\bea
\ln\frac{\mu^2}{m_S^2} +i\frac{\pi}{2} =
\frac{1}{2}\ln\frac{\mu^2}{-m_S^2-i\epsilon}+\frac{1}{2}\ln\frac{\mu^2}{m_S^2}\, ,
\eea
so logarithms of both $+m_S^2$ and $-m_S^2$ appear. This is because there are corrections
coming from soft exchanges between the two initial state particles, similar to Drell-Yan, which give rise to 
logs of $-m_S^2$, and soft exchanges between initial and final state particles, similar to deep inelastic scatttering, which give rise to logs of $+m_S^2$. From Eq.~(\ref{gammaH}) we can infer the  double logarithms in the $O(\alpha_s)$ 
corrections to $C_i(\mu),~i=S,P$,
\bea
C_i(\mu) = C_i^{(0)}\left[1- \frac{C_A \alpha_s}{4 \pi}\left(\frac{1}{2}\log^2\left(\frac{-m_S^2-i \epsilon}{\mu^2}\right) + \frac{1}{2}\log^2\left(\frac{m_S^2}{\mu^2}\right) + ...\right) \right] \, ,
\eea
where $...$ denotes terms without double logs. 
From this we see that if $\mu=m_S$,  $C_i(m_S)$ gets a $\pi^2$-enhanced contribution:
$C_i(m_S) = C_i^{(0)}(1+ C_A \alpha_s (m_S) \pi/8)$. For the range of $m_S$ considered 
in this paper, $\alpha_s(m_S) \leq 0.1$.
If $\alpha_s=0.1$, this $\pi^2$-enhanced correction
increases the cross section by about 24\%, and is half as big as the corresponding $\pi^2$-enhanced
contribution to Higgs production. Refs.~\cite{Ahrens:2008nc,Ahrens:2008qu} argued that the $\pi^2$-enhanced terms dominate the fixed-order corrections to Higgs production, and that these  terms can be resummed to all orders by evolving the hard function from the 
scale $m_H^2$ to the scale $-m_H^2$. They also showed that the leading terms exponentiate.
In our case, setting $\mu^2_f = -m_S^2$ does not  remove the factor 
of $\pi^2$ in the hard coefficient. The double logs vanish if
\bea
\mu^2 =e^{\pi (\pm1-i)/2} m_S^2,
\eea
but it is not clear that evolving to this complex scale will give a sensible resummation the $\pi^2$-enhanced contribution.
 Below we will  calculate the $K$-factor with the next-to-leading order (NLO) $\pi^2$-enhanced correction.
Even if the $\pi^2$-enhanced corrections exponentiate, the  NLO correction should be a good 
approximation to the resummed result since $1+C_A\alpha_s \pi/4$ and
$\exp(C_A\alpha_S \pi/4)$ differ by less than 3\% for $\alpha_s=0.1$.

\begin{figure}[t]
\begin{center}
\epsfig{file=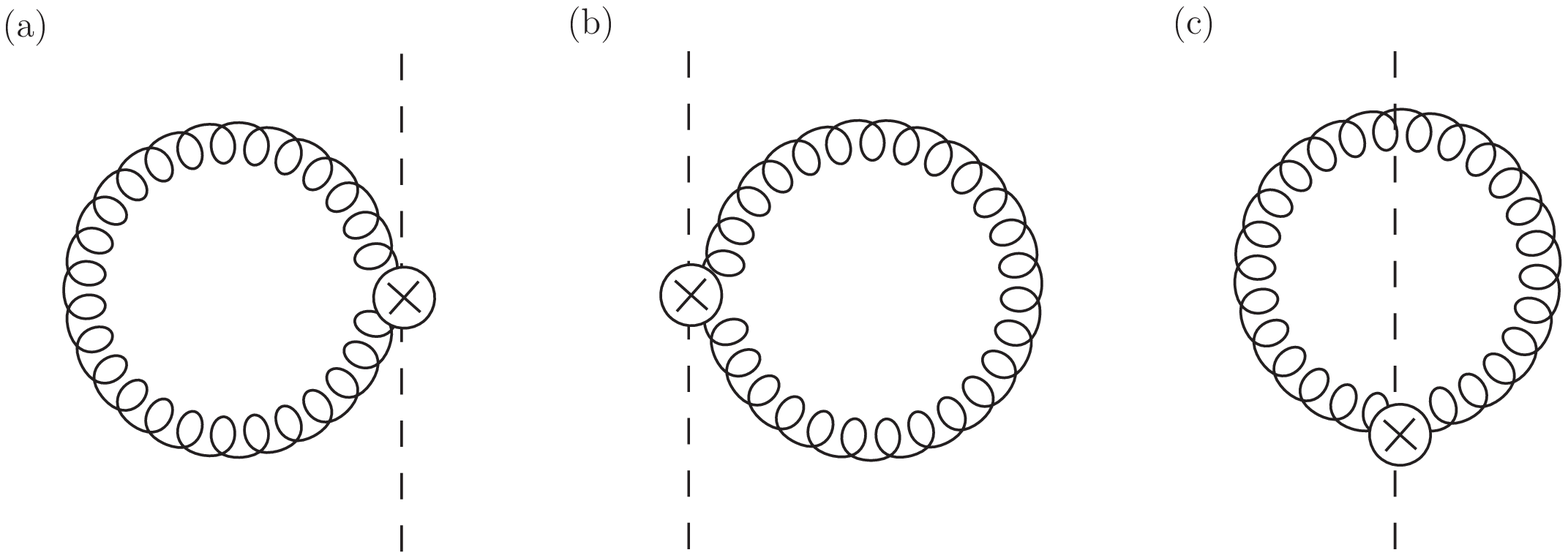, width=13.5cm, height=4.5cm}
\end{center}
\vspace{-0.7cm}
\caption{\baselineskip 3.0ex  
 One loop corrections to the soft function. The dashed line represents the cut. The diagram (a) and its Hermitian conjugate (b) describe the virtual soft gluon radiation and the diagram (c) denotes real soft gluon radiation.  
} 
\label{fig2} 
\end{figure}

The soft functions in Eqs.~(\ref{softS}) and (\ref{softP}) can be computed perturbatively 
when $\mu_s  \sim m_S (1-z) \gg \Lambda_{\rm QCD}$. The  Feynman diagrams in Fig.~\ref{fig2} give us the $O(\alpha_s)$ corrections to $\bar{S}_{S} (m_S(1-z),\mu)$ and $\bar{S}_{P}(m_S(1-z),\mu)$:
\begin{eqnarray} 
\label{soft1l}
&&\bar{S}_S^{(1)} (m_S(1-z),\mu) = \bar{S}_P^{(1)} (m_S(1-z),\mu) \\
&&~~~=\frac{\alpha_s}{\pi} N_c \Biggl\{\delta(1-z) \Bigl[\frac{1}{\veps_{\mathrm{UV}}^2} +\frac{1}{\veps_{\mathrm{UV}}} \Bigl(\frac{1}{2}+\ln \frac{\mu^2}{m_S^2}\Bigr) 
+1-\frac{\pi^2}{4}+\frac{1}{2}\ln\frac{\mu^2}{m_S^2}+\frac{1}{2}\ln^2\frac{\mu^2}{m_S^2}\Bigr] \nonumber \\ 
&&~~~-\Bigl(\frac{2}{\veps_{\mathrm{UV}}}+1+2\ln \frac{\mu^2}{m_S^2} \Bigr)\frac{1}{(1-z)_+}+4\Bigl(\frac{\ln(1-z)}{1-z}\Bigr)_+ \Biggr\}, \nonumber 
\end{eqnarray}
where the plus distributions are defined in the standard way.
%\begin{eqnarray} 
%\label{plus1}
%\int^1_0 dz \frac{f(z)}{(1-z)_+} &=& \int^1_0 dz \frac{f(z)-f(1)}{1-z}, \\
%\label{plus2} 
%\int^1_0 dz \Bigl(\frac{\ln(1-z)}{1-z}\Bigr)_+ f(z) &=& \int^1_0 dz \frac{\ln(1-z)}{1-z} \Bigl(f(z) -f(1)\Bigr). 
%\end{eqnarray} 
%So, up to next-to-leading order (NLO), we can identify $S_f$ and $S_d$, which help us simplify the factorization formular in Eq.~(\ref{scaf1}).  
Note that there is no IR divergence in the sum of the real and virtual diagrams in Fig.~\ref{fig2} \cite{KLN}. The IR finiteness of the soft function can be easily understood in SCET because the soft function is just the Wilson coefficient obtained at the second-step matching. 

To obtain the resummed cross section we employ the method of 
momentum space resummation developed in Refs.~\cite{Becher:2006nr,Becher:2006mr,Becher:2007ty,Ahrens:2008nc,Ahrens:2008qu}. The resummed result can be written as
\begin{equation} 
\label{scar} 
\sigma (pp\to S_i X) = \tau \int^1_z \frac{dz}{z} V_i(z,m_S,\mu_f) F(\tau/z, \mu_f),~~~i=S,P
\end{equation} 
where the resummation function $V_i (z,m_S,\mu_f)$ is given by
\begin{equation} 
\label{Vr}
V_{S,P}(z,m_S,\mu_f)=H_{S,P} (m_S,\mu_h) U(\mu_h,\mu_s,\mu_f) \frac{z^{-\eta}}{(1-z)^{1-2\eta}} 
\tilde{S}_{S,P}(\partial_{\eta},\mu_s) \frac{e^{-2\gamma_E \eta}}{\Gamma (2\eta)}.
\end{equation} 
Here 
%$H(m_S, \mu_h)$ is the hard-coefficient function, $\propto |C_i(m_s, \mu_h)|^2$,
$\tilde S_{S,P}(\partial_\eta,\mu_s)$ is defined in terms of the Laplace transform of the soft functions,
and the evolution function $U(\mu_h,\mu_s,\mu_f)$ is a product of terms obtained 
from evolving the hard function to the scale $\mu_h$ and 
the Laplace transform soft function to the scale $\mu_s$. To NLL accuracy, we find 
\begin{equation} 
\label{Us}
\ln U (\mu_h,\mu_s,\mu_f) = \ln \Bigl[4 SU_{\mathrm{NLL}}(m_S,\mu_s) + \frac{B_1^S}{\beta_0} \ln \frac{\alpha_s (\mu_s)}{\alpha_s (m_S)} + \frac{B_1^g}{\beta_0} \ln \frac{\alpha_s (\mu_f)}{\alpha_s (\mu_s)}\Bigr], 
\end{equation}
where $SU_{\mathrm{NLL}}$ is 
\begin{equation} 
SU_{\mathrm{NLL}}(\mu_1,\mu_2) = \frac{A_1}{4\beta_0^2} \Bigl[
\frac{4\pi}{\alpha_s(\mu_1)} \Bigl(1-\frac{1}{r} -\ln r \Bigr)+\Bigl(\frac{A_2}{A_1} -\frac{\beta_1}{\beta_0} \Bigr)(1-r+\ln r) +\frac{\beta_1}{2\beta_0} \ln^2 r \Bigr],
\end{equation} 
where $r=\alpha_s (\mu_2)/\alpha_s (\mu_1)$. The parameters $A_1$, $A_2$, $B_1^S$, 
and $B_1^g$ appear in the anomalous dimensions of the hard and soft functions, and 
are given in the Appendix. The parameter $\eta$ is defined in terms
of an integral over the cusp anomalous dimension (see Ref.~\cite{Becher:2006nr})
 and in our calculation $\eta = (A_1/\beta_0) \ln (\alpha_s (\mu_f)/\alpha_s (\mu_s))$.
 In our  case,  $\eta <0$ since $\mu_s < \mu_f$ and hence the integral in 
Eq.~(\ref{scar}) is singular. The integral is then defined by analytic  continuation from positive $\eta$. 
We will choose $\mu_h=\mu_f=m_S$. 
For this choice of $\mu_h$ there are no large logs of $m_S^2/\mu^2$ in $H_{S,P}(m_S,\mu_h)$
and $H_{S,P}(m_S,\mu_h) = H_{S,P}^{(0)} (m_S,\mu_h)$ to NLL accuracy. 
In order to resum logarithms of $1-z$ we should choose the
scale $\mu_s= m_S(1-z)$, however, this will lead to divergences in the $z$ integral as
 the running coupling will cross the Landau pole as $z\to1$. Practically, it is simpler to choose $\mu_s$ to be a scale parametrically smaller than $\mu_h$.

\begin{figure}[t]
\begin{center}
\epsfig{file=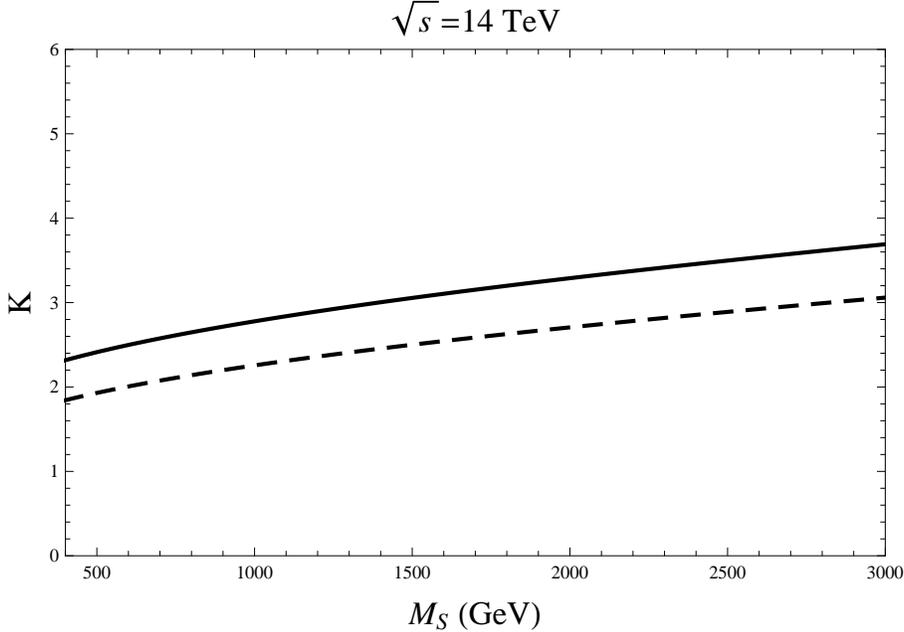, width=12cm, height=9cm}
\end{center}
\vspace{-0.7cm}
\caption{\baselineskip 3.0ex  
$K$-factor for the single color-octet production at LHC where $\sqrt{s} = 14$ TeV. The straight   (dashed) line denotes NLL evaluation with (without) $\pi^2$-evolution.   
} 
\label{fig3} 
\end{figure}

\begin{figure}[t]
\begin{center}
\epsfig{file=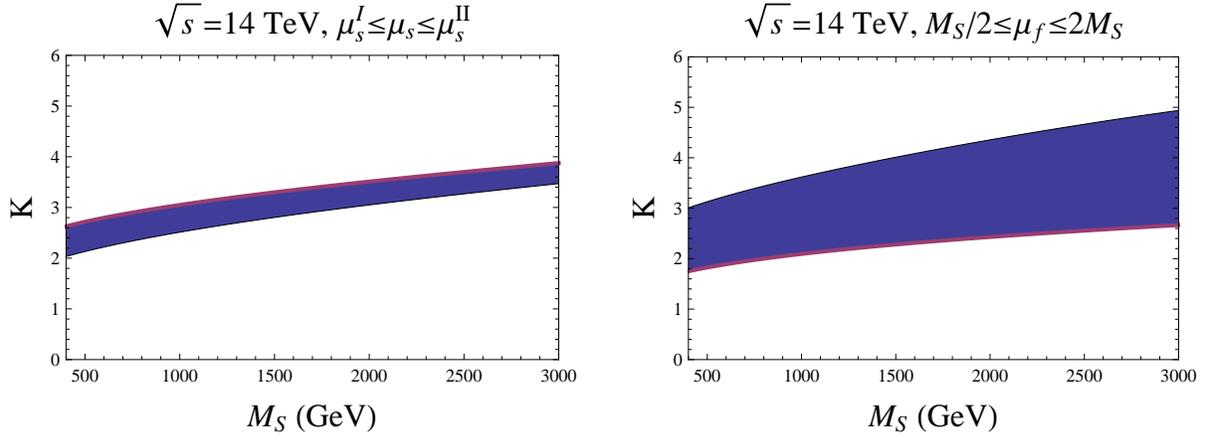, width=16cm, height=6.4cm}
\end{center}
\vspace{-1cm}
\caption{\baselineskip 3.0ex  
Scale dependences of the $K$-factor. 
} 
\label{fig4} 
\end{figure}

We first present our results in terms of a $K$-factor, defined as the ratio of leading order and 
NLL resummed cross sections, which is given by
\begin{equation} 
\label{Kf} 
K_{S,P}(m_S^2, \tau) = \int^1_z \frac{dz}{z} V_{S,P}(z,m_S,\mu_f) F(\tau/z, \mu_f) \Biggl/ \Bigl(H_{S,P}^{(0)}(m_S,\mu_f)\int^1_z \frac{dz}{z} F_{\mathrm{LO}} (\tau/z, \mu_f)\Bigr),
\end{equation} 
where $F_{\rm{LO}}$ is a convolution of PDFs at LO. This result is universal in that it is independent of
the NP model.  $O(\alpha_S(m_S))$ corrections to the hard coefficient
can depend on the NP model but this beyond the accuracy we are working. 
For our numerical calcuations, we use the LO $\alpha_s$, setting $\alpha_s (M_Z) = 0.1205$ and
 $m_t = 170.9$ GeV. For the gluon PDF's  we use CTEQ5 at NLO \cite{Lai:1999wy}.
In order to determine $\mu_s$, we follow the procedure of Ref.~\cite{Becher:2007ty} and
calculate  the convolution of the one-loop expression for $\bar{S}_i(m_S(1-z),\mu)$ in Eq.~(\ref{soft1l}) with $F(\tau/z,\mu)$ in Eq.~(\ref{sca}).
The scale $\mu_s$ is chosen so that the higher order corrections to the soft function are under perturbative control. This is accomplished by specifying $\mu_s^I$ and $\mu_s^{II}$: The scale $\mu_s^I$ is defined by starting with $\mu_s=\mu_h$ and lowering 
$\mu_s$ until the $O(\alpha_S)$ correction is less than 15\%. The scale $\mu_s^{II}$ is chosen so that the one-loop correction is minimized. We use the average $(\mu_s^I +\mu_s^{II})/2$
in Fig.~\ref{fig3}. The solid line is the result for the $K$-factor with the $\pi^2$-enhanced correction included. The $K$-factor varies from about 2.4 for $m_S = 500$ GeV to about 3.6 for $m_S= 3$ TeV.
As expected, the resummation of threshold corrections significantly enhances the cross section
and becomes more important as $m_S$ increases. The dashed line in Fig.~\ref{fig3} is the result
without the $\pi^2$-enhanced correction. This correction increases the $K$-factor by ~25\% and is independent of $m_S$.
In Fig.~\ref{fig4}, we show the variation
in the  prediction as $\mu_s$ is varied between $\mu_s^I$ and $\mu_s^{II}$. The uncertainty
from the choice of $\mu_s$ is $\pm 15\%$ for $m_S = 500$ GeV, and decreases with increasing
$m_S$. The variation with the choice of $\mu_f$ is also shown in Fig.~\ref{fig4}. The sensitivity to the choice of $\mu_f$ is greater and introduces an uncertainty of $\pm 25\%$. The dependence 
on the scales $\mu_s$ and $\mu_f$ should decrease when higher order corrections are included.

\begin{figure}[t]
\begin{center}
\epsfig{file=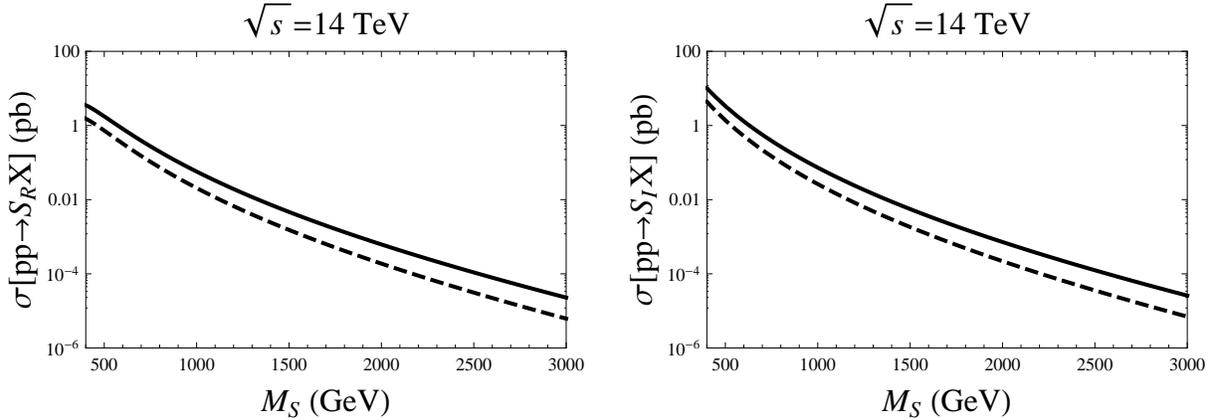, width=16cm, height=6.4cm}
\end{center}
\vspace{-1cm}
\caption{\baselineskip 3.0ex  
The scattering cross sections employing Manohar-Wise Model at the LHC. In the both plots, the straight (dashed) lines denote the results at NLL (LO). 
} 
\label{fig5} 
\end{figure}

In Fig.~\ref{fig5}, we show our calculation of the color-octet scalar production cross section in the Manohar-Wise Model \cite{Manohar:2006ga}. In this model, the two real components of the 
complex  color-octet scalars are denoted $S^0_R$ and $S^0_I$, where $S^0_R$ is a scalar and 
$S^0_I$ is a pseudoscalar if $\eta_U$ is chosen to be real.
The LO calculation of their production cross sections from  Ref.~\cite{Gresham:2007ri}
are the dashed lines in Fig.~\ref{fig4}, and our NLL results are the solid lines. At $m_S = 1$ TeV, we obtain  $\sigma_{\rm{NLL}} (pp \to S_R X) = 57$ fb and $\sigma_{\rm{NLL}} (pp \to S_I X) = 73$ fb. 
We have fixed the parameters $\eta_U=1$ and  $\lambda_{4,5}=1$ as in Ref.~\cite{Gresham:2007ri}. The NLL results are almost 3 times as large as the LO results, which are $\sigma_{\rm{LO}} (pp \to S_R X) = 21$ fb and $\sigma_{\rm{LO}} (pp \to S_I X) = 26$ fb. 

In summary we have used SCET to derive a factorization theorem for  color-octet scalar  production at the LHC. The factorization theorem can be used to resum large threshold corrections which  have a significant impact on the total cross section. It is universal 
in the sense that all details dependent on NP models are encoded in the Wilson coefficients. The factorization theorem is similar to Higgs production, however, some details are different because the 
final state particle is colored. Because there are both soft exchanges between initial state partons as well as between partons in the initial and final states, the structure of double logarithms and corresponding $\pi^2$-enhanced corrections is different. We obtained a resummed calculation
of $\sigma (pp\to SX)$ to NLL accuracy. The resummed cross sections are 2-4 times larger than the LO cross section, depending on the mass of the color-octet scalar. Uncertainties from varying $\mu_S$ and $\mu_f$ in these calculations are $\pm 15\%$ and $\pm25\%$, respectively. NNLL log resummation and higher order perturbative corrections will be required to reduce scale dependence of the resummed cross section. Further development of the factorization theorem to account for scales besides 
$m_S$ and $m_S(1-z)$ maybe required if color-octet scalars are actually discovered. For example,
precision measurements of the mass may require taking into account the width of the color-octet scalar, 
as is required for determining the top quark mass~\cite{Fleming:2007qr}. 
 
\acknowledgments 

This work was supported in part by the U.S. Department of Energy under 
grant numbers DE-FG02-05ER41368 and DE-FG02-05ER41376.

\appendix

\section{Large N resummation in moment space}
Here we present our results for the resummed cross section in moment space. All the results below are taken in the large $N$ limit. For the renormalized soft function $S_N$ we find 
\begin{equation} 
\label{SHmoments} 
\bar{S}^N_{S,P} (\mu)= \int^1_0 z^{N-1} \bar{S}_{S,P}(m_S(1-z),\mu) = 1+\frac{\alpha_s}{\pi} N_c \Bigl(\frac{1}{2} \ln^2 \frac{\mu^2\bar{N}^2}{m_S^2}+\frac{1}{2} \ln \frac{\mu^2\bar{N}^2}{m_S^2} +1+ \frac{\pi^2}{12} \Bigr) + \mO (\alpha_s^2), 
\end{equation} 
where $\bar{N} = N e^{\gamma_E}$. From Eq.~(\ref{SHmoments}), we notice that the choice
 $\mu = m_S/\bar{N}$ minimizes the large logarithms. The $\mu$-independence of cross section
 implies the following RGE:
\begin{equation} 
\label{SHRGE} 
\mu \frac{d}{d\mu} \bar{S}^N_i(\mu) = \Bigl(2\gamma_g^N -2 \mathrm{Re}[\gamma_i]\Bigr) \bar{S}_i^N (\mu),~~~i=S,P,
\end{equation} 
where $\gamma_g^N$ is the well-known Altarell-Parisi evolution kernel for the gluon PDF in the moment space,
\begin{equation} 
\label{ggn}
\gamma_g^N = \frac{\alpha_s}{\pi} C_A \Bigl[2\ln \bar{N} - \Bigl(\frac{11}{6}-\frac{n_f}{9}\Bigr) \Bigr] +\mO(\alpha_s^2).
\end{equation} 
From our results, Eqs.~(\ref{gammaH}), (\ref{SHmoments}), and (\ref{ggn}) we can easily see that Eq.~(\ref{SHRGE}) is satisfied to first order in $\alpha_s$.
 If we take the moments of $\sigma (pp\to S_i X)$ in Eq.~(\ref{factthm}), the result is
 \begin{equation} 
\label{sigN}
\sigma_N(pp \to S_i X) = \int^1_0 d\tau \tau^{N-1} \sigma (pp\to S_i X) = H_i(m_S, \mu_s) \bar{S}_i^N  (\mu_s, \mu_f) [f_{g/p}^N (\mu_f)]^2 +O\left(\frac{1}{N}\right),
\end{equation} 
where we identified $\mu_h = m_S$. Here we employed the two-step matching: the hard coefficient $H_{S,P} (m_S)$ at the scale $m_S$ is evolved down to the soft scale $\mu_s$ and then the soft function $\bar{S}(\mu_s)$ obtained at $\mu_s$ can be evolved to the factorization scale $\mu_f$. This is equivalent to the scaling evolution realized in Eq.~(\ref{Vr}), where the hard and soft function are evolved from $\mu_h$ and $\mu_s$ to $\mu_f$ respectively, but the soft function's renormalization behavior compensates the evolution of the hard function from $\mu_s$ to $\mu_f$. 
So the exponentiated matching coefficients $C_{S,P}$ and $\bar{S}_{S,P}^N$ are given by
\begin{eqnarray} 
C_{S,P}(m_S,\mu_s) &=& C_{S,P}(m_S) e^{-I_1 (m_S, \mu_s)} = C_{S,P}(m_S) \exp \Bigl[-\int^{m_S}_{\mu_s} \frac{d\mu}{\mu} \gamma_{S,P} (\mu) \Bigr],
\\
\bar{S}_{S,P}^N (\mu_s,\mu_f) &=& \bar{S}_{S,P}^N  (\mu_s) e^{-I_2 (\mu_s,\mu_f)} = \bar{S}_{S,P}^N  (\mu_s) \exp \Bigl[-2\int^{\mu_s}_{\mu_f} \frac{d\mu}{\mu} \gamma_g^N (\mu) \Bigr].
\end{eqnarray}

Therefore Eq.~(\ref{sigN}) can be rewritten as
\begin{eqnarray} 
\label{sigN1} 
\sigma_N (pp\to S_iX)&=& H_i (m_S)~e^{-2 \mathrm{Re}[I_1 (m_S, \mu_s)]} \bar{S}_i^N (\mu_s) e^{-I_2 (\mu_s,\mu_f)} [f_{g/p}^N (\mu_f)]^2 \\
\label{sigN2} 
&=& H_i (m_S) \exp{[\mathcal{G} (m_S,\mu_f)]} [f_{g/p}^N (\mu_f)]^2, 
\end{eqnarray} 
where we set $\mu_s = m_S/\bar{N}$ in the second equality, and then 
the exponential factor $\mathcal{G}_S$ can be expanded as  
\begin{equation} 
\mathcal{G} (m_S,\mu_f) = \ln \bar{N} g_S^{(0)} + g_S^{(1)} (m_S,\mu_f) + \alpha_s (m_S) g_S^{(2)} (m_S,\mu_f)+ \cdots. 
\end{equation} 
Here each of the coefficients $g_S^{(i)},~i=0,1,2$ are correspond to the resummed results at LL, NLL, and NNLL accuracies, respectively. 

For the computation of the exponentiation factor up to NLL accuracy, we need to expand $\mathrm{Re}[\gamma_{S,P}]$ and $\gamma_g^N$ up to second order in $\alpha_s$ 
\begin{eqnarray} 
\label{anomalexpH} 
\mathrm{Re}[\gamma_{S}] &=&\mathrm{Re}[\gamma_{P}]= -\frac{\alpha_s}{4\pi} \Bigl[ A_1 \ln \frac{\mu^2}{m_S^2} + B_1^S\Bigr] -\Bigl(\frac{\alpha_s}{4\pi}\Bigr)^2 A_2 \ln \frac{\mu^2}{m_S^2} +\mO(\alpha_s^2), \\
\label{anomalexpg}
\gamma_g^N &=& \frac{\alpha_s}{4\pi} \Bigl[ A_1 \ln  \bar{N}^2 - B_1^g\Bigr] +\Bigl(\frac{\alpha_s}{4\pi}\Bigr)^2 A_2 \ln \bar{N}^2 +\mO(\alpha_s^2), 
\end{eqnarray} 
where $\alpha_s \ln (\mu/m_S) \sim \alpha_s \ln \bar{N}$ are treated as $\mO (1)$, and the coefficients of the large logarithms, $A_i$ denote the coefficients of the cusp anomalous dimension \cite{Korchemsky:1992xv}. In the above equations, $A_1$, $B_1^S$, and $B_1^g$ were already given in Eqs.~(\ref{gammaH}) and (\ref{ggn}), and $A_2$ is  $8N_c [(67/18 - \pi^2/6)N_c -5 n_f/9]$ \cite{Korchemsky:1992xv}. 

After a brief calculation using Eqs.~(\ref{anomalexpH}) and (\ref{anomalexpg}), we find 
\begin{eqnarray} 
\label{gH0}
g_S^{(0)} &=& \frac{A_1}{\lambda \beta_0} \Bigl[2\lambda +(1-2\lambda)\ln(1-2\lambda)\Bigr], \\
\label{gH1} 
g_S^{(1)} &=& \frac{1}{\beta_0} \Bigl[(B_1^g-B_1^S)\ln(1-2\lambda)+2A_1\lambda  \ln\frac{\mu_f^2}{m_S^2} \Bigr] -\frac{A_2}{\beta_0^2} \Bigl[2\lambda+\ln(1-2\lambda)\Bigr]\\
&&+\frac{\beta_1 A_1}{2\beta_0^2} \Bigl[4\lambda + (2+\ln(1-2\lambda))\ln(1-2\lambda)\Bigr],\nonumber 
\end{eqnarray} 
where $\lambda=\alpha_s \beta_0\ln \bar{N} /(4\pi)$, and $\beta_{0,1}$ are the first two coefficients of the QCD $\beta$ function.
Here note that we have set $\mu_f \sim m_S$. However, if we choose $\mu_f$ as $\mu_f \le \mu_s \sim m_S/\bar{N}$, the logarithm $\ln(\mu_f /m_S)$ should be power-counted as $\mO (\ln \bar{N})$.

\end{document}